# Direct and Inverse Magnetocaloric effects in A-site ordered $PrBaMn_2O_6$ manganite in low magnetic fields

A.M. Aliev*[,1], A.G. Gamzatov, A.B. Batdalov, V.S. Kalitka[2], A.R. Kaul[2]

[1]Amirkhanov Institute of Physics, Daghestan Scientific Center of RAS, 367003, Makhachkala, Russia

[2]Moscow State Univercity, 119899, Moscow, Russia

*lowtemp@mail.ru

## Abstract

The magnetocaloric effect (MCE) of A-site ordered $PrBaMn_2O_6$ manganite has been studied by direct methods and by the specific heat measurements. Direct measurements of the MCE in low magnetic fields were performed using recently proposed modulation technique and by classic direct method in high fields. Direct and inverse MCE are observed at Curie and Neel points correspondingly. A value of the inverse MCE in the heating run is less than in the cooling regime. This effect can be attributing to competition between ferromagnetic and antiferromagnetic interactions. Indirectly estimated and direct MCE values considerably differ in around first order AF transition.

Recently, half-doped manganite perovskites with the A-site order $RBaMn_2O_6$ have been attracting of interest due to their novel physical properties [1-4]. $RBaMn_2O_6$ displays remarkable features: the charge/orbital order (CO) transition at relatively high temperatures, a new stacking variation of the CE-type CO with a fourfold periodicity along the c-axis, the presence of structural transition and electronic phase segregation and these features strongly depend on an ordering degree of R and Ba cations [5]. But a nature of appearance of the cation-ordered state in Ba-substituted manganites is not understood entirely so far.

The ground state of $PrBaMn_2O_6$ is a coexistence of ferromagnetic metal (with $T_C$=310-320 K) and A-type antiferromagnetic metal (with $T_N$=200-270 K) [5]. Such spread of the critical temperatures follows from a substantial dependence of the physical properties of these materials on synthesis conditions [6, 7].

* Corresponding author: e-mail lowtemp@mail.ru

The ceramic sample $PrBaMn_2O_6$ was prepared by chemical homogenization method. The X-ray diffraction analysis confirms that the sample is single-phase and the unit cell parameters are a=3.9007(1), c=7.7486(4) (*P4mm* group).

Commonly the isothermal magnetic entropy change of magnetic materials is estimated indirectly from magnetic measurements and the use of the Maxwell relation (MR),

$$\Delta S_M(T,H) = \int_0^H \left(\frac{\partial M}{\partial T}\right) dH \qquad (1)$$

But the use of Eq. (1) is only valid if the system is in thermodynamic equilibrium. If there occurs a first-order magnetic phase transition, magnetization values that do not correspond to the equilibrium value can be measured. So, there is an approximation when Eq. (1) is used to estimate $\Delta S_M$ from magnetization data for first-order phase transition system, since the possible metastable nature of the measured state is not taken into consideration. The overestimation of $\Delta S$ from using the Maxwell relation in nonequilibrium system can be as high as 1/3 of the value obtained under equilibrium [8], and even ten or more times greater [9].

A correct value of the entropy change can be determined either by integrating the MR only within the transition region [9] to avoid overestimations caused by ferromagnetic portions or by using the Clausius–Clapeyron (CC) equation $\Delta S = -\frac{\Delta M}{(\Delta T/\Delta H)}$ [9, 10].

The estimation of MCE from specific heat using equation

$$\Delta S = \int_0^T \frac{C_{p,0} - C_{p,H}}{T} dT \qquad (2)$$

at magnetostructural phase transitions, can result to ambiguous values of MCE as well. Even though in real materials at I-order phase transitions an infinite specific heat does not observe, the use of this relation has some limitations because it needs to know a temperature dependence of the heat capacity starting from 0 degree of Kelvin. In consideration of a weak dependence of the heat capacity on a magnetic field at I-order transition and a sensitivity of the integrating procedure to value of integrated quantity, this can result to poor accuracy of the MCE estimation. So, it can be concluded the direct measurements are most preferable and reliable techniques for study of magnetocaloric properties of materials at first order magnetostructural/charge ordered transitions.

The MCE measurements in low fields were carried out by a modulation technique [11]. The essence of the method consists in low frequency alternating magnetic field influence on a sample and register of a.c. temperature response. The measurements are carried out at frequency of 0.3 Hz, at different amplitudes of the magnetic field. The specific heat was measured by the a.c. -calorimeter method.

The temperature dependence of magnetization of $PrBaMn_2O_6$ at heating is shown in Fig. 1. On the base of MR (see insert in Fig. 1), we can expect direct and inverse MCE at FM and AFM magnetostructural phase transitions with the nearly equal values.

The temperature dependence of the specific heat of $PrBaMn_2O_6$ in 80-350 K temperature range is shown in Fig. 2. A high temperature anomaly of the specific heat with maximum at $T$= 304.1 K is caused by a ferromagnetic phase transition. A decreasing of the temperature results in the second anomaly of the specific heat, with peaks at 214.7 K at cooling and at 243.5 K at heating. The wide hysteresis ($\Delta T \approx 30$ K) indicates the first order nature of the transition and points out on significant structure change at the antiferromagnetic transition. Magnetic field $H$=11 kOe smoothes the high temperature specific heat anomaly and shifts its maximum to higher temperature ($T_{Cp\ peak}$ = 313.4 K). Only weak magnetic field effect on antiferromagnetic anomaly of the specific heat is observed. Determination of the MCE using Eq. (2) reveals direct and inverse MCE in the studied system (insert in Fig.2).

In Fig. 3 the temperature dependences of MCE in $PrBaMn_2O_6$ obtained by a modulation technique near ferromagnetic phase transition are shown. The maxima of the temperature change on all curves are near $T$=308 K and amount to $\Delta T$ = 0.065 K at amplitude of field change $\Delta T$ = 750 Oe. To extrapolate low field MCE values to high fields, we have measured field dependence of the magnetocaloric effect at 308 K (Insert in Fig. 3). The field dependence of MCE in a ferromagnet near $T_C$ can be expressed as $\Delta S = aH^{2/3}$, where $n$=2/3, $a$ is a constant [12]. From this follows that $\Delta T \approx H^{2/3}$ in the vicinity of $T_C$. Note, in the latter expression a field dependence of the specific heat is not taken into account. Fitting our experimental results give $n$=0.90, and it results to $\Delta T$=0.67 K at magnetic field change of 1.1 T. It is the mean value of the MCE in manganites [13]. Though MCE in $PrBaMn_2O_6$ does not achieve large values, the transition width is more than 60 K even in low fields. It is important the effect is observed in a temperature interval of 260-320 K, what is optimal region for room temperature magnetocaloric materials.

More complex behavior of the MCE is observed at the antiferromagnetic transition (Fig. 4). First, the inverse magnetocaloric effect is observed, what additionally confirms the antiferromagnetic nature of the transition. As the specific heat, the temperature dependence of MCE is characterized by hysteresis also. MCE value, obtained at heating, is less than that obtained at cooling. A competition between the antiferromagnetic and ferromagnetic ordering is suggested to be the reason for this. The total value of the MCE can be presented as the sum $\Delta T_{tot} = \Delta T_F + \Delta T_{AF}$, where $\Delta T_F$ and $\Delta T_{AF}$ are magnetocaloric effects due to ferromagnetic and antiferromagnetic paraprocesses correspondingly. Near Neel point at cooling regime, $\Delta T_{tot} = \Delta T_{AF}$, because $\Delta T_F = 0$ far from Curie point in low fields. At heating, total MCE $\Delta T_{tot}$ is the sum of $\Delta T_{AF}$ with negative sign and nonzero $\Delta T_F$ with positive sign because the Curie point is considerably close to the Neel point.

It results to $\Delta T_{\text{tot, cooling}} > \Delta T_{\text{tot, heating}}$. With magnetic field increasing $\Delta T_F$ will increase steadily, whereas $\Delta T_{AF}$ will increase until $H<H_{cr}$ ($H_{cr}$ is critical magnetic field that induce AFM→FM transition).

The same effect must be observed around the ferromagnetic transition but only slightly above the antiferromagnetic transition, since the antiferromagnetic ordering will fast decrease above $T_N$. And the picture is really observed above $T_N$, and the MCE curves quickly merge at approximation to $T_C$.

The MCE at high fields, obtained by classic direct method at heating, are shown in Fig. 5. The peak value of MCE ($\Delta T$=0.528 K at $\Delta H$=11 kOe) at FM transition is less than obtained by extrapolation of modulation technique results ($\Delta T$=0.67 K at $\Delta H$=11 kOe). By-turn, MCE from specific heat data ($\Delta T$=0.79 K at $\Delta H$=11 kOe) is greater only on 15% in comparison with the modulation technique results. More discrepancy between the data is observed at FM-AFM transition. At low fields the inverse MCE is observed. With field raising the compensation temperature shifts to low temperatures and AFM-FM crossover is takes place. And at field change of 11 kOe we observe direct MCE ($\Delta T$=0.13 K) around $C_p$ anomaly. In contrast to this, from the specific heat data inverse MCE value follows ($\Delta T$=-0.20 K). Thus, the significant differences of MCE values around I–order phase transition can be attributed to utilized techniques. Our direct measurements show that MCE around I-order magnetostructural transition have ordinary nature and does not achieve giant values. Early reported giant MCE values around first-order charge/orbital ordering transition in manganites and other materials [14 -17] can be attributed to inadequate use of MR. In this work we show that use of Eq. (2) for MCE estimation at first-order transition can result to significant inaccuracy. It can conclude from comparison of results obtained by different methods that modulation technique has advantages as compared with other ones: high sensitivity, ability of measure of MCE in very low fields of minute samples, high rate of measurements and etc.

**Acknowledgements** This work was partly supported by RFBR (Grant Number 09-08-96533) and the Physics Department of RAS.

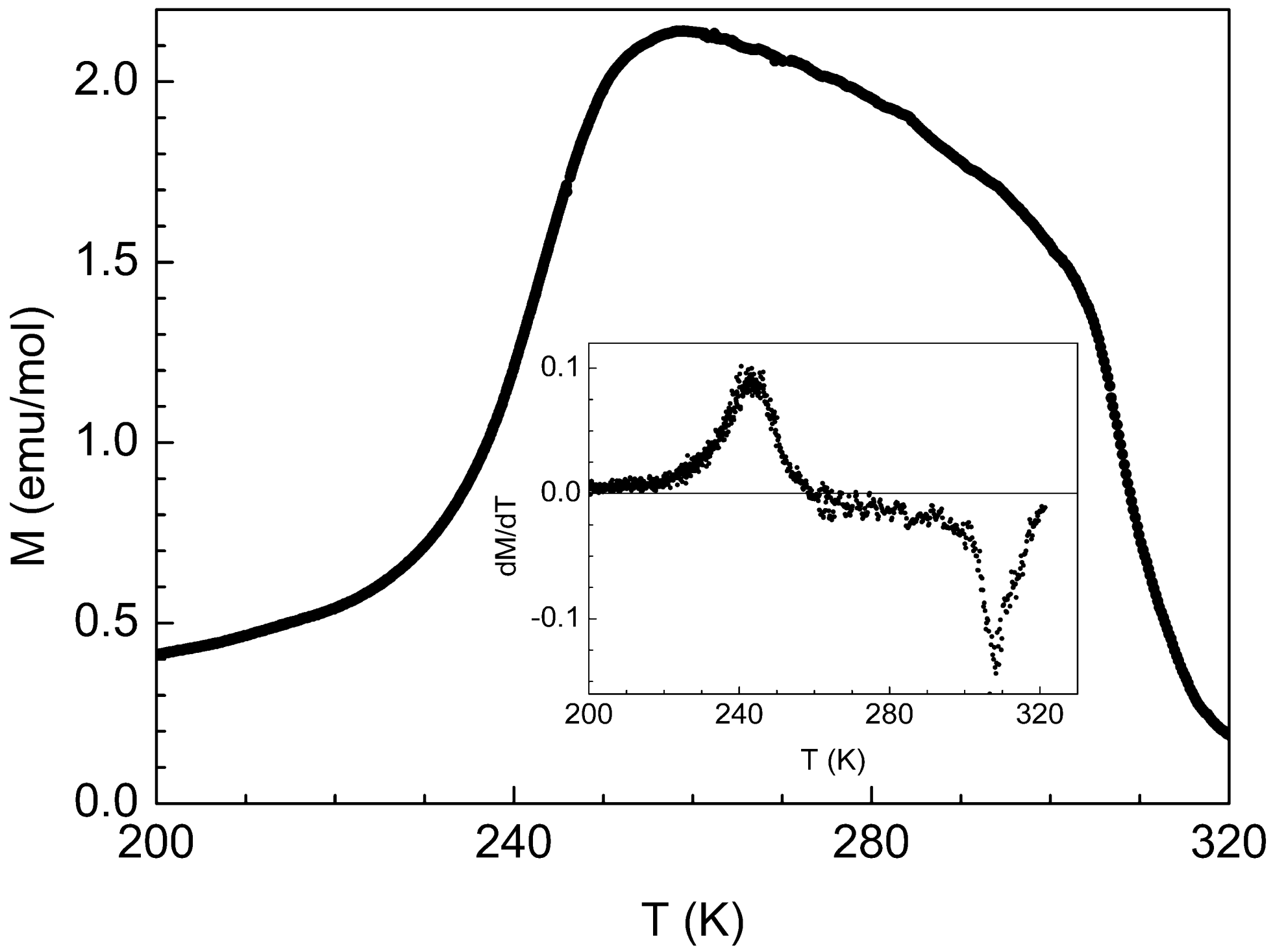

**Figure 1.** Temperature dependences of the magnetization at heating. Insert – dM/dT(T).

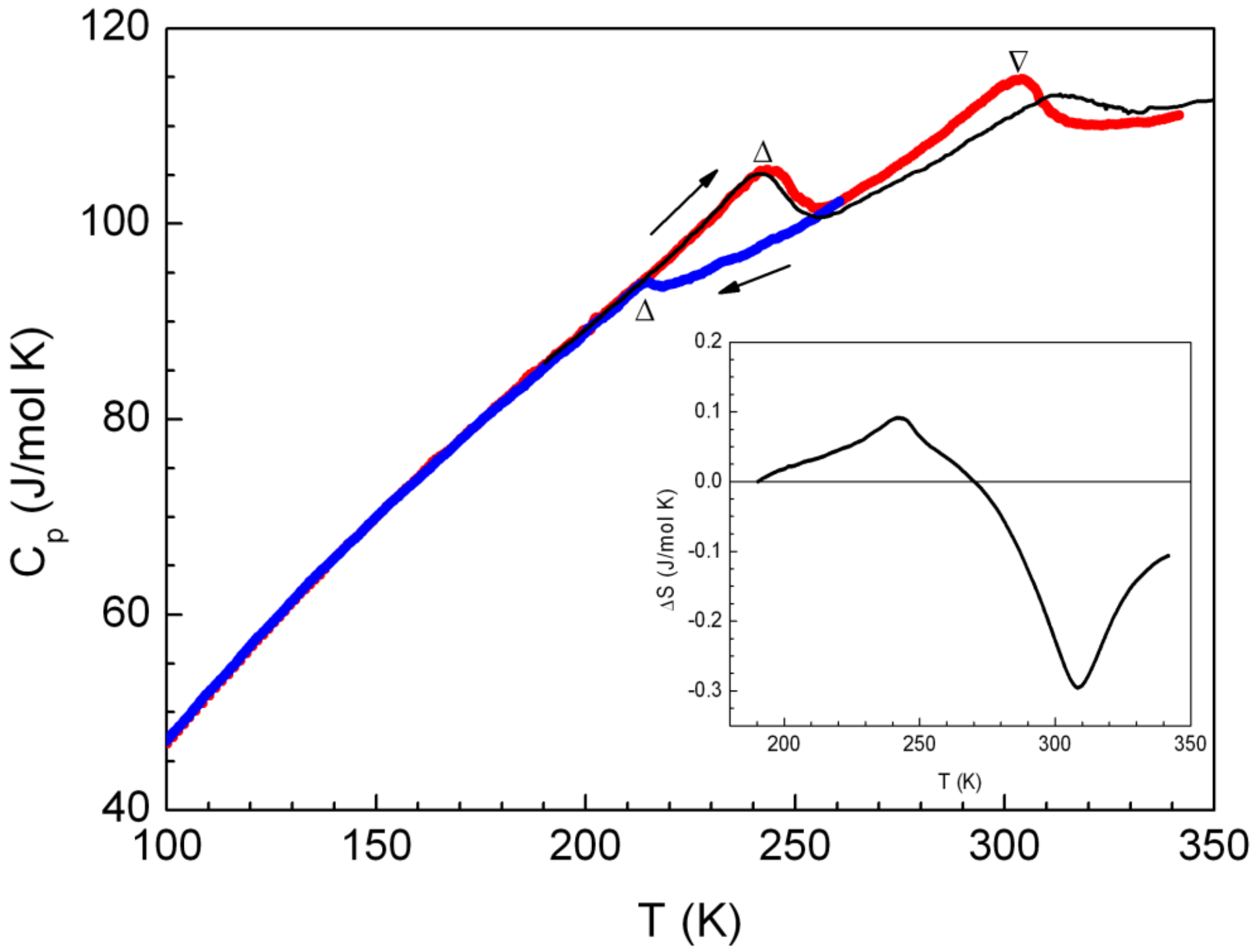

**Figure 2.** Temperature dependences of the specific heat of $PrBaMn_2O_6$ at cooling and heating. Up triangles denote Neel points, down triangle – Curie point. Black line – $C_p$ at $H$=11 kOe. In insert –Δ$S$($T$) curve at Δ$H$ = 11 kOe, estimated from specific heat data.

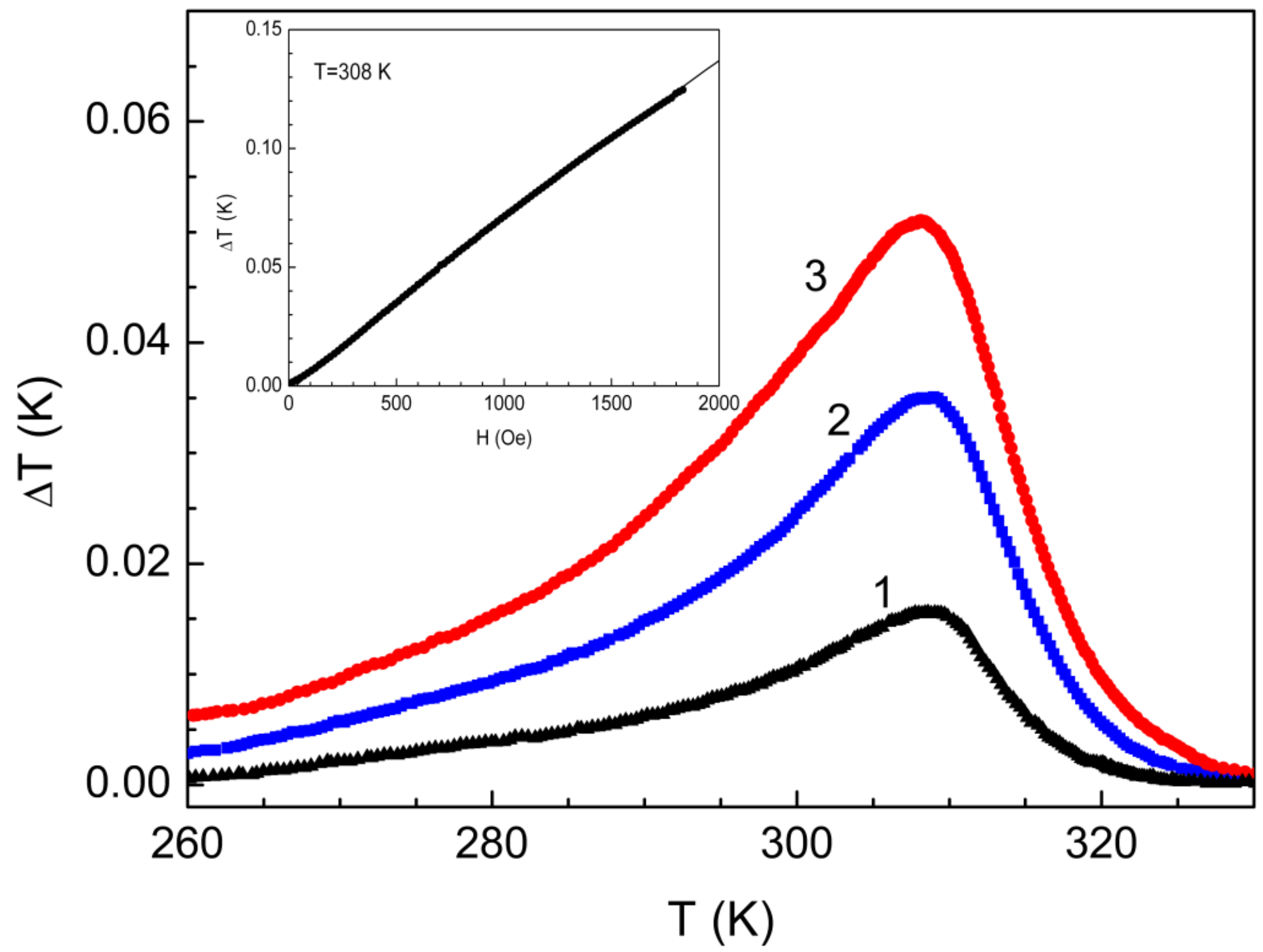

**Figure 3.** MCE in $PrBaMn_2O_6$ around ferromagnetic phase transition (modulation technique). 1 – Δ*H*=250 Oe, 2 - Δ*H*=500 Oe, 3 - Δ*H*=750 Oe. Insert – field dependence of MCE at *T*=308 K.

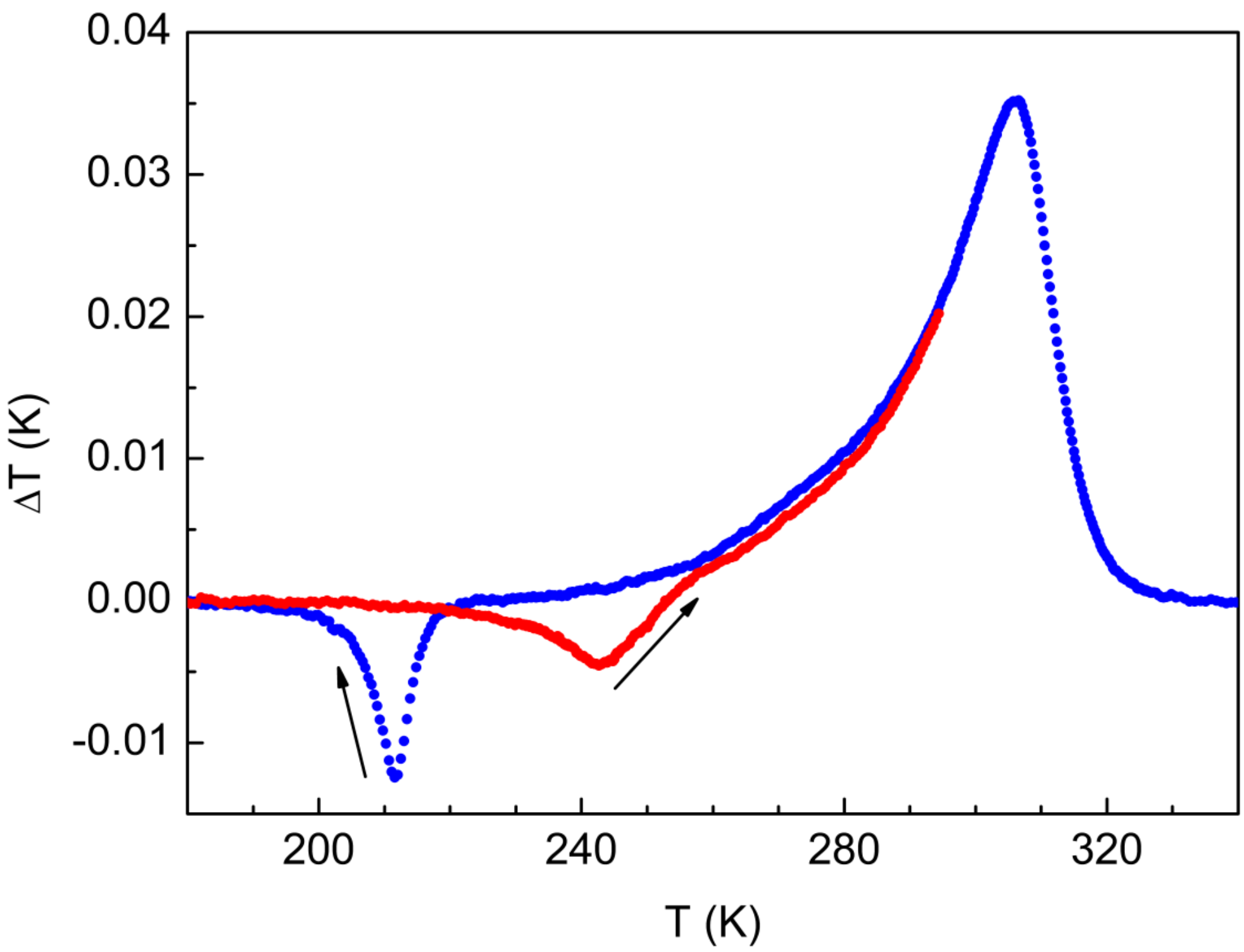

**Figure 4.** Temperature dependence of the MCE of $PrBaMn_2O_6$ in cooling and heating runs at magnetic field change Δ*H*=500 Oe (modulation technique).

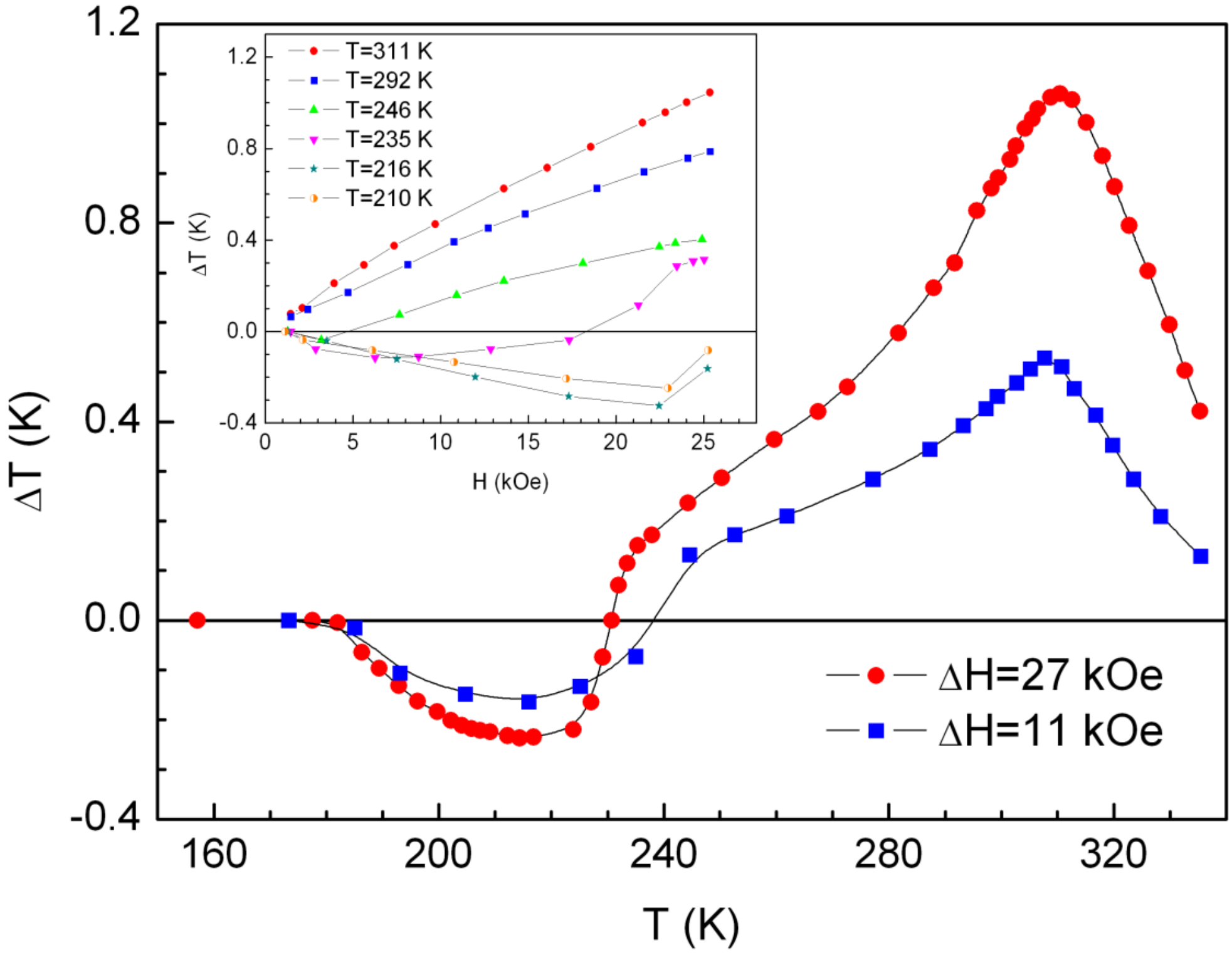

**Figure 5.** Temperature dependence of the MCE of $PrBaMn_2O_6$ at heating (classic direct method). Lines are guide to eyes. In insert – field dependence of MCE at various temperatures.